\documentclass{llncs}
\usepackage{makeidx}  % allows for indexgeneration
\usepackage[tight,footnotesize]{subfigure}
\usepackage{cite}
\usepackage{url}
\usepackage{tikz}
\usetikzlibrary{positioning}
\usepackage{verbatim}
\usepackage{booktabs}
\usepackage{multirow}
\usepackage{listings}
\usepackage{float}
\usepackage{etoolbox}
\begin{document}
\frontmatter          % for the preliminaries
\pagestyle{headings}  % switches on printing of running heads
\addtocmark{Hamiltonian Mechanics} % additional mark in the TOC

\mainmatter              % start of the contributions
\title{Streaming Applications on Heterogeneous Platforms}
\titlerunning{Hamiltonian Mechanics}  % abbreviated title (for running head)
%\author{Zhaokui Li\inst{1} \and Jianbin Fang}
\author{Zhaokui Li,  Jianbin Fang, Tao Tang, Xuhao Chen \and Canqun Yang}
\authorrunning{Zhaokui Li et al.} % abbreviated author list (for running head)
\tocauthor{Ivar Ekeland, Roger Temam, Jeffrey Dean, David Grove,
Craig Chambers, Kim B. Bruce, and Elisa Bertino}
\institute{Software Institute, College of Computer, \\ National University of Defense Technology, Changsha, China\\
\email{zhaokuili@yeah.net, \{j.fang, tangtao84, chenxuhao, canqun\}@nudt.edu.cn}
}

\maketitle              % typeset the title of the contribution

\begin{abstract}
%Using multiple streams can improve the overall system performance by mitigating the data transfer overhead on heterogeneous systems. However, not all applications are suitable for multiple streams. In this work, we use a total of 53 benchmarks which are from the Rodinia Benchmark Suite, the Parboil Benchmark Suite, the NVIDIA SDK, and the AMD APP SDK, but almost half of these benchmarks is not suitable for streaming. So we present a set of method to prejudge whether the application is overlappable or suitable for streamlization, then we classify the streamlization application into three patterns, different categories, different processing methods. To conclude, our work provides one mechanism to judge whether one application is worthwhile to streamlizing and one guideline about how to streamlize the code. 

Using multiple streams can improve the overall system performance by mitigating the data transfer overhead on heterogeneous systems. Currently, very few cases have been streamed to demonstrate the streaming performance impact and a systematic investigation of \textit{streaming necessity} and \textit{how-to} over a large number of test cases remains a gap. In this paper, we use a total of 56 benchmarks to build a statistical view of the data transfer overhead, and give an in-depth analysis of the impacting factors. Among the heterogeneous codes, we identify two types of \textit{non-streamable} codes and three types of \textit{streamable} codes, for which a streaming approach has been proposed. Our experimental results on the CPU-MIC platform show that, with multiple streams,  we can improve the application performance by up 90\%. Our work can serve as a generic flow of using multiple streams on heterogeneous platforms.

\keywords{Multiple Streams, Heterogeneous Platforms, Performance.}
\end{abstract}

\section{Introduction}

%\textbf{Motivation}

%\textbf{Research Questions:} (1) whether a given application is worthy to use streaming mechanisms?  (2) how can we streamlize an application? 

%\textbf{Contribution:} (1) we give a statistical view of xxx, (2) we illustrate how to streamlize different types of applications. 

Heterogeneous platforms are increasingly popular in many application domains~\cite{citeulike:2767438}. The combination of using a host CPU combined with a specialized processing unit (e.g., GPGPUs or Intel Xeon Phi) has been shown in many cases to improve the performance of an application by significant amounts. Typically, the host part of a heterogeneous platform manages the execution context while the time-consuming code piece is offloaded to the coprocessor. Leveraging such platforms can not only enable the achievement of  high peak performance, but increase the \textit{performance per Watt ratio}.

% Introducing multiple streams
Given a heterogeneous platform, how to realize its performance potentials  remains a challenging issue. In particular, programmers need to explicitly move data between host and device over \texttt{PCIe} before and/or after running kernels. The overhead counts when data transferring takes a decent amount of time, and determines whether to perform offloading is worthwhile~\cite{citeulike:13920330, citeulike:6102210, citeulike:13920339}. To hide this overhead, overlapping kernel executions with data movements is required. To this end, \textit{multiple streams} (or \textit{streaming mechanism}) has been introduced, e.g., CUDA Streams~\cite{tr:cuda:best}, OpenCL Command Queues~\cite{website:opencl_ref}, and Intel's hStreams~\cite{tr:hstreams:arch}. These implementations of \textit{multiple streams} spawn more than one streams/pipelines so that the data movement stage of one pipeline overlaps the kernel execution stage of another~\footnote{In the context, the streaming mechanism is synonymous with \textit{multiple streams}, and thus we refer the \textit{streamed code} as \textit{code with multiple streams}.}.

%  Prior work 
Prior works on multiple streams mainly focus on GPUs and the potential of using multiple streams on GPUs is shown to be significant~\cite{citeulike:13920334, citeulike:9715521, citeulike:13920353, DBLP:journals/ieicet/InoNH13}.  Liu et al. give a detailed study into how to achieve optimal task partition within an analytical framework for AMD GPUs and NVIDIA GPUs~\cite{citeulike:13920353}. In~\cite{citeulike:9715521}, the authors model the performance of asynchronous data transfers of CUDA streams to determine the optimal number of streams. However, these studies have shown very limited number of cases, which leaves two questions unanswered: (1) whether each application is required and worthwhile to use multiple streams on a given heterogeneous platform?, (2) whether each potential application is \textit{streamable} or \textit{overlappable}? If so, how can we stream the code? 

% Our focus 
To systematically answer these questions, we (1) build a statistical view of the data transfer (\texttt{H2D} and \texttt{D2H}) fraction for a large number of test cases, and (2) present our approach to stream different applications. Specifically, we statistically show that more than 50\% test cases (among 223) are not worthwhile to use multiple streams. The fraction of \texttt{H2D} varies over platforms, applications, code variants, and input configurations. Further, we identify two types of \textit{non-streamable} codes (\texttt{Iterative} and \texttt{SYNC}) and three categories of \textit{streamable} code based on task dependency ( \texttt{embarrassingly independent}, \texttt{false dependent}, and \texttt{true dependent}). Different approaches are proposed to either eliminate or respect the data dependency. As case studies, we stream 13 benchmarks of different categories and show their streaming performance impact. Our experimental results show that using multiple streams gives a performance improvement ranging from 8\% to 90\%.  To the best of our knowledge, this is the first comprehensive and systematic study of multiple streams in terms of both \textit{streaming necessity} and \textit{how-to}. To summarize, we make the following contributions: 

\begin{itemize}

\item We build a statistical view of the data transfer fraction ($R$) with a large number of test cases and analyze its impacting factors (Section~\ref{sec:stat}). 

\item We categorize the heterogeneous codes based on task dependency and present our approach to stream three types of applications (Section~\ref{sec:approach}). 

\item We show a generic flow of using multiple streams for a given application: calculating $R$ and performing code streaming (Section~\ref{sec:stat},~\ref{sec:approach}). 
%: calculating the \texttt{H2D} faction, categorizing the application, and performing streamlization. 

\item We demonstrate the performance impact of using multiple streams on the CPU-MIC platform with 13 streamed benchmarks (Section~\ref{sec:exp}). 

\end{itemize}

\section{Related Work} \label{sec:related_work}
%\section{Related Work} \label{sec:related_work}
In this section, we list the related work on pipelining, multi-tasking, workload partitioning, multi-stream modeling, and offloading necessity. 
%In this section, we discuss the related work on \textit{pipelining}, \textit{multi-tasking}, and \textit{optimization space pruning}. 
%Topic 1: Traditional pipelining
%We hereby discuss the related work. 

\textbf{Pipelinining}  is widely used in modern computer architectures~\cite{citeulike:1102804}. Specifically, the pipeline stages of an instruction run on different functional units, e.g., arithmetic units or data loading units. In this way, the stages from different instructions can occupy the same functional unit in different time steps, thus improving the overall system throughput. Likewise, the execution of a heterogeneous application is divided into stages (\texttt{H2D}, \texttt{KEX}, \texttt{D2H}), and can exploit the idea of software pipelining on the heterogeneous platforms. 

% Topic 2: Multiple streams
%Many studies of GPU computing applications focus on individual kernel performance and ignore that many applications have to regularly transfer data over the \texttt{PCIe} bus. \cite{xx}\cite{xx} show that PCIe transfers can have a large impact on performance for many applications.  

%Topic 3: Multi-tasking
\textbf{Multi-tasking} provides concurrent execution of multiple applications on a single device. In~\cite{citeulike:11069829}, the authors propose and make the case for a GPU multitasking technique called \textit{spatial multitasking}. The experimental results show that the proposed spatial multitasking can obtain a higher performance over cooperative multitasking. In~\cite{WendeSteinkeCordes2014}, Wende et al. investigate the concurrent kernel execution mechanism that enables multiple small kernels to run concurrently on the Kepler GPUs. Also, the authors evaluate the Xeon Phi offload models with multi-threaded and multi-process host applications with concurrent coprocessor offloading~\cite{citeulike:13920403}. Both multitasking and multiple streams share the idea of spatial resource sharing. Different from multi-tasking, using multiple streams needs to partition the workload of a single application (rather than multiple applications) into many tasks. 

%In~\cite{xx}, Miyamoto et al. propose a dynamic offload scheduler which assigns processor resources of Phi to tasks by an offload model. They also show their effectiveness of their approach.

%Kepler GPUs introduce the Hyper-Q feature with 32 hardware managed work queues for concurrent kernel execution. 

%Topic 3: Multi-device programming
%Topic 4: workload partition
\textbf{Workload Partitioning: }  There is a large body of workload partitioning techniques, which intelligently partition the workload between a CPU and a coprocessor at the level of algorithm~\cite{citeulike:13920419}\cite{citeulike:13920423} or during program execution~\cite{citeulike:13920424}\cite{citeulike:10176926}. Partitioning workloads aims to use unique architectural strength of processing units and improve resource utilization~\cite{citeulike:13920415}. In this work, we focus on how to efficiently utilize the coprocessing device with multiple streams. Ultimately, we need to leverage both workload partitioning and multiple streams to minimize the end-to-end execution time. 

\textbf{Multiple Streams Modeling:} In~\cite{citeulike:9715521}, Gomez-Luna et al. present performance models for asynchronous data transfers on different GPU architectures. The models permit programmers to estimate the optimal number of streams in which the computation on the GPU should be broken up. In~\cite{citeulike:13920334}, Werkhoven et al. present an analytical performance model to indicate when to apply which overlapping method on GPUs. The evaluation results show that the performance model are capable of correctly classifying the relative performance of the different implementations. In~\cite{citeulike:13920353}, Liu et al. carry out a systematic investigation into task partitioning to achieve maximum performance gain for AMD and NVIDIA GPUs. Unlike these works, we aim to evaluate the necessity of using multiple streams and investigate how to use streams systematically. Using a model on Phi to determine the number of streams will be investigated as our future work.
%Unlike these works, we discuss the heuristics of reducing the search space when determining the factors. Using a model on Phi will be investigated as our future work. 

\textbf{Offloading Necessity: } Meswani et al. have developed a framework for predicting the performance of applications executing on accelerators~\cite{citeulike:14070672}. Using automatically extracted application signatures and a machine profile based on benchmarks, they aim to predict the application running time before the application is ported. Evaluating offloading necessity is a former step of applying multiple streams. In this work, we evaluate the necessity of using multiple streams with a statistical approach. 

\section{A Statistical View} \label{sec:stat}
In this section, we give a statistical view of how many applications are worthwhile to be streamed on heterogeneous platforms, and analyze the factors that impact the streaming necessity. 

\subsection{Benchmarks and Datasets}
As shown in Table~\ref{tbl:benchmark_intro}, we use a large number of benchmarks that cover a broad range of interesting applications domains for heterogeneous computing. These benchmarks are from the Rodinia Benchmark Suite, the Parboil Benchmark Suite, the NVIDIA SDK, and the AMD APP SDK. In total, we employ 56 benchmarks and 223 configurations. The details about how applications are configured are summarized in Table~\ref{tbl:benchmark_intro}. Note that we remove the redundant applications among the four benchmark suites when necessary.

\begin{table}[!t]
\caption{Applications, Inputs and Configurations}
\begin{center}
\scalebox{0.70}{
\begin{tabular}{|c|c|c|c|c|}
\hline 
\textbf{Suite} & \textbf{Applications} & \textbf{Input}  & \textbf{Applications} & \textbf{Input} \\ \hline
Rodinia (18)  & 
\begin{tabular}[c]{@{}c@{}}
	backprop\\
	b+tree \\
	dwt2d,gaussian, lud \\
	hearwall, lavaMD, leukocyte \\
	kmeans \\
	pathfinder \\
\end{tabular}  & 
\begin{tabular}[c]{@{}c@{}}
	$10 \times \{2^{16}, 2^{17}, 2^{18}, 2^{19}, 2^{20}\}$\\ 
	Kernel1, Kernel2 \\
	$ 2^{10}, 2^{11}, 2^{12}, 2^{13}, 2^{14}$\\ 
	10, 20, 30, 40, 50 \\
	$\{1, 3, 10, 30, 100\} \times 100000$ \\
	$(\{1, 2, 4\}\times 10^{5} , \{100, 200, 400\})$
 \end{tabular} & 
\begin{tabular}[c]{@{}c@{}} 
	bfs \\ 
	cfd  \\
	myocyte, srad\\
	hotspot \\
	nn,  nw\\
	streamcluster \\
\end{tabular} & 
\begin{tabular}[c]{@{}c@{}} 
	graph\{512K, 1M, 2M, 4M, 8M\} \\ 
	0.97K, 193K, 0.2M \\
	100, 200, 300, 400, 500\\ 
	$ 2^{9}, 2^{10}, 2^{11}, 2^{12}, 2^{13}$\\ 
	$ 2^{10}, 2^{11}, 2^{12}, 2^{13}, 2^{14}$\\ 
	$ 100 \times \{2^{10}, 2^{11}, 2^{12}, 2^{13}, 2^{14}$\}\\ 
\end{tabular}  \\ \hline

Parboil (9)  & 
\begin{tabular}[c]{@{}c@{}} 
	spmv\\ 
	mri-gridding \\ 
	tpacf \\
	sgemm \\
	lbm \\
\end{tabular}  & 
\begin{tabular}[c]{@{}c@{}}
	small, medium, large\\ 
	small \\
	small, medium, large \\
	small, medium \\
	short, long \\
\end{tabular}  &  
\begin{tabular}[c]{@{}c@{}} 
	stencil\\ 
	cutcp \\
	bfs \\
	mri-q \\ 
\end{tabular}  & 
\begin{tabular}[c]{@{}c@{}}
	small, default\\ 
	small, large \\
	1M, NY, SF, UT \\
	small, large \\
\end{tabular}  \\ \hline

NVIDIA SDK (17)  & 
\begin{tabular}[c]{@{}c@{}}BlackScholes\\ DCT8x8 \\ DXTCompression \\ Histogram \\ MatVecMul \\ Reduction \\ Transpose \\ VectorAdd \end{tabular}  & 
\begin{tabular}[c]{@{}c@{}} $10^6 \times \{4, 8, 12, 16, 20\}$\\ $2^{10} \times \{1, 2, 3, 4, 8\}$ \\ lena \\ 1, 2, 3, 4, 5 \\ 256, 128, 64, 32, 16 \\ $2^{10} \times 10^3 \times  \{1, 2, 3, 4, 8\}$ \\ $2^{10} \times \{1, 2, 3, 4, 8\}$ \\ $2^{10} \times \{1, 2, 4, 8, 16\}$  \end{tabular}  & 
\begin{tabular}[c]{@{}c@{}}ConvolutionSeparable\\ DotProduct  \\ FDTD3d \\ MatrixMul \\ QuansiRandomGenerator \\ Reduction-2 \\ Tridiagonal \\ FastWalshTransform \\ ConvolutionFFT2D \end{tabular} &  
\begin{tabular}[c]{@{}c@{}} $2^{10} \times \{1, 2, 3, 4, 8\}$\\ $2^{10} \times 10^3 \times  \{1, 2, 3, 4, 8\}$ \\ 10, 20,30, 40, 50 \\ 6, 7, 8, 9, 10 \\ $2^{10} \times 10^3 \times  \{1, 2, 3, 4, 8\}$ \\ $2^{10} \times 10^3 \times  \{1, 2, 3, 4, 8\}$ \\ 32, 64, 128, 256, 512 \\ 8M \\ 62500k \end{tabular} \\ \hline

AMD SDK (12) & 
\begin{tabular}[c]{@{}c@{}}BinomialOption\\ BoxFilter \\ FloydWarshall \\ RadixSort \\ ScanLargeArrays \\ URNG \end{tabular}  & 
\begin{tabular}[c]{@{}c@{}} $2^{10} \times \{1, 2, 4, 8, 16\}$ \\  BoxFilter\_Input \\ $2^{10} \times \{1, 2, 3, 4, 5\}$ \\ $2^{12} \times \{12, 13, 14, 15, 16\}$ \\ $2^{10} \times \{1, 2, 4, 8, 16\}$ \\ 1, 2, 3, 4, 5 \end{tabular} & 
\begin{tabular}[c]{@{}c@{}} BitonicSort\\ DwtHaar1D  \\ MonteCarloAsian \\ RecursiveGaussian \\ StringSearch \\ PrefixSum \end{tabular} &  
\begin{tabular}[c]{@{}c@{}}$2^{20} \times \{1, 2, 4, 8, 16\}$ \\ $2^{10} \times 10^3 \times  \{1, 2, 3, 4, 8\}$ \\ $2^{10} \times \{1, 2, 3, 4, 5\}$  \\ default  \\ 1, 2, 3, 4, 5 \\ 1024k  \end{tabular} \\ \hline

\end{tabular}}
\end{center}
\label{tbl:benchmark_intro}
\end{table}

\subsection{Experimental Platforms} \label{subsec:exp:platform}
The heterogeneous platform used in this work includes a dual-socket Intel Xeon CPU (12 cores for each socket) and an Intel Xeon 31SP Phi (57 cores for each card). The host CPUs and the cards are connected by a \texttt{PCIe} connection. As for the software, the host CPU runs Redhat Linux v7.0 (the kernel version is 3.10.0-123.el7.x86\_64), while the coprocessor runs a customized uOS (v2.6.38.8). Intel's MPSS (v3.6) is used as the driver and the communication backbone between the host and the coprocessor.  Also, we use Intel's multi-stream implementation~ \texttt{hStreams (v3.5.2)} and Intel's OpenCL SDK (v14.2).  Note that the applications in Table~\ref{tbl:benchmark_intro} are in OpenCL, while the pipelined versions are in \texttt{hStreams}. 

\subsection{Measurement Methodology}
A typical heterogeneous code has three parts: (1) transferring data from host to device (\texttt{H2D}), (2) kernel execution (\texttt{KEX}), and (3) moving data from device back (\texttt{D2H}). To measure the percentage of each stage, we run the codes in a strictly stage-by-stage manner.  Moreover, we perform 11 runs and calculate the median value. Before uploading datasets, buffer allocation on the device side is required. Due to the usage of the \textit{lazy allocation policy}, the allocation overhead is often counted into \texttt{H2D}. Thus, we argue that \texttt{H2D} might be larger than the actual host-to-device data transferring time. 

\subsection{Results and Analysis}
\begin{figure}[!h]
\centering
\includegraphics[width=0.60\textwidth]{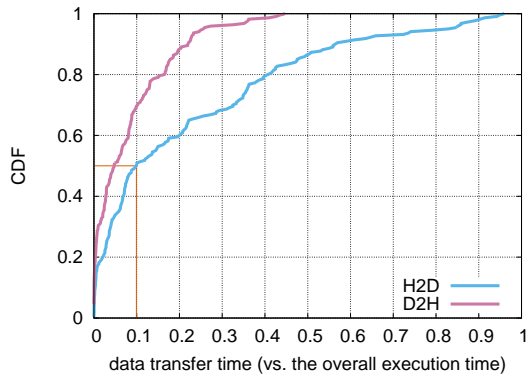}
\caption{The CDF curve for data transfers between the host and the accelerator.}
\label{fig:cdf}
\end{figure}

We define~\textit{data transfer ratio ($R$)} as the fraction of the data transfer time to the total execution time, and take this metric ($R$) as an indicator of whether it is necessary to use multiple streams. Figure~\ref{fig:cdf} shows the CDF distribution of the \texttt{H2D} and \texttt{D2H} duration versus the overall execution time ($R_{H2D}$ and $R_{D2H}$). We observe that the CDF is over 50\% when $R_{H2D}=0.1$. That is, the \texttt{H2D} transfer time takes less than 10\% for more than 50\% configurations. Meanwhile, the number is even larger (around 70\%) for the \texttt{D2H} part. In the remaining contents, we will focus on $R_{H2D}$ and use $R$ (instead of $R_{H2D}$) for clarity. 

\subsubsection{The Impact of Input Datasets}
\begin{figure}[!h]
\centering
\subfigure[lbm.]{\label{fig:stat:lbm}\includegraphics[width=0.35\textwidth]{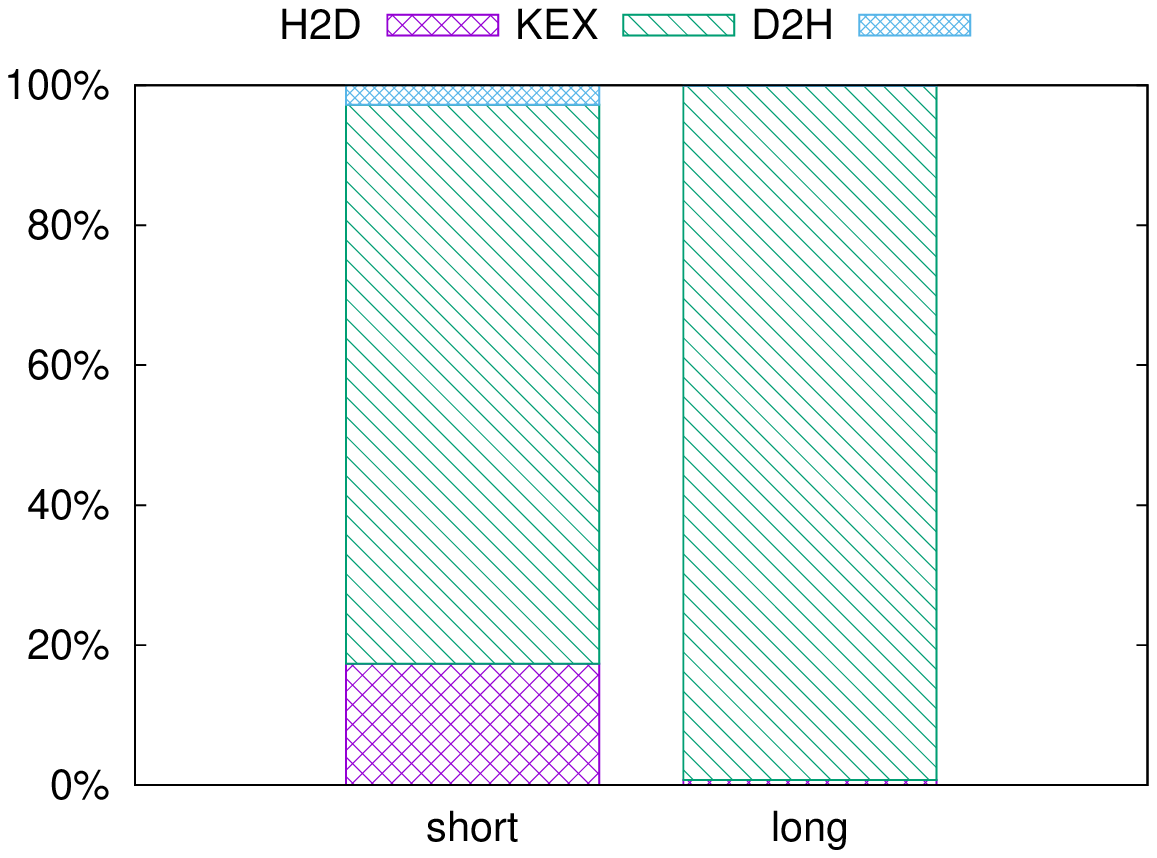}}
\subfigure[FDTD3d.]{\label{fig:stat:fdtd}\includegraphics[width=0.35\textwidth]{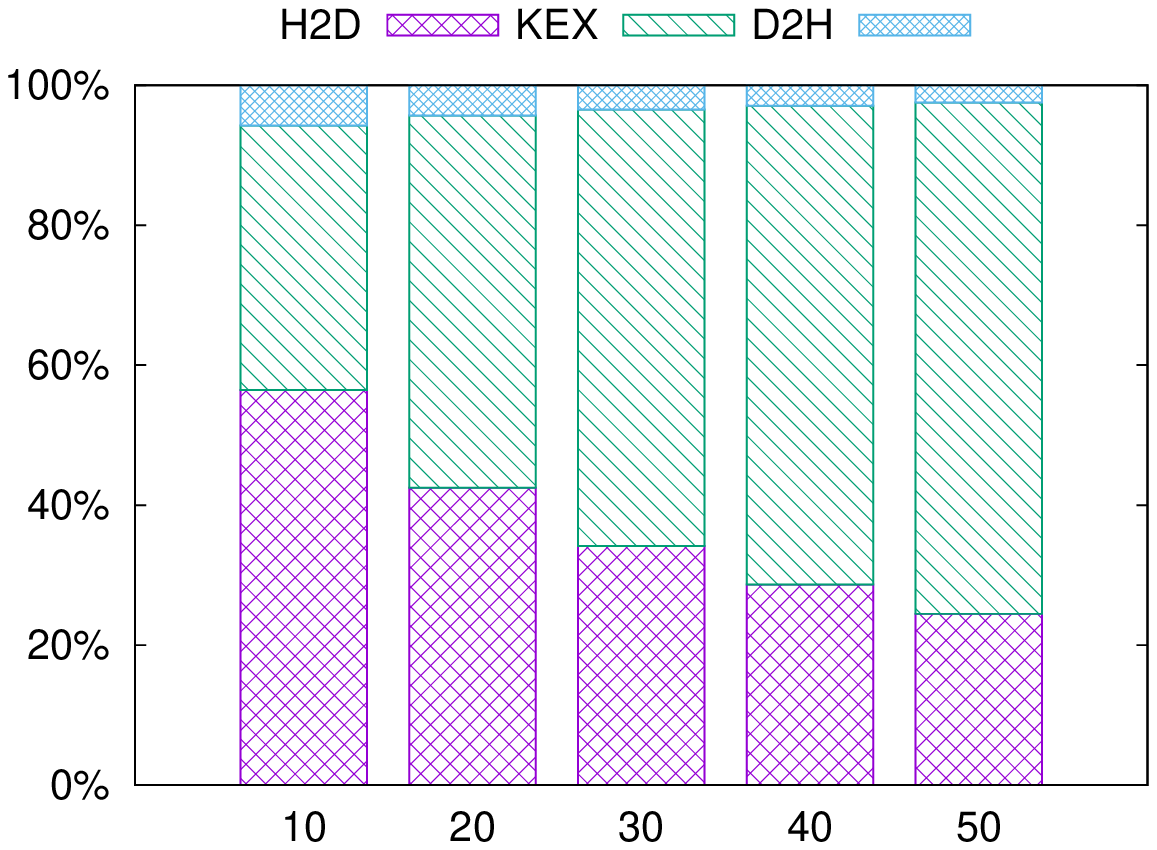}}
\caption{\texttt{R} changes over datasets for \texttt{lbm} and \texttt{FDTD3d}. }
\label{fig:stat_datasets}
\end{figure}

Typically, the \texttt{H2D} ratio  will remain when changing the input datasets. This is because the computation often changes linearly over the data amount. But this is not necessarily the case. Figure~\ref{fig:stat_datasets} shows how $R$ changes with the input datasets for \texttt{lbm} and \texttt{FDTD3d}, respectively. We note that, for \texttt{lbm}, using the \texttt{short} configuration takes a decent amount of time to move data from host to device, while the data amount takes a much smaller proportion for the \texttt{long} configuration. For \texttt{FDTD3d}, users have to specify the number of time steps according to their needs.  We note that the kernel execution time increases over time steps. When streaming such applications, it is necessary to focus on the commonly used datasets. 

%the value of pipelining also varies over the input datasets. 

%When increasing the number of time steps, the kernel execution takes longer.

\subsubsection{The Impact of Code Variants}
\begin{figure}[!h]
\centering
\subfigure[reduction v1.]{\label{fig:stat:reductionv1}\includegraphics[width=0.35\textwidth]{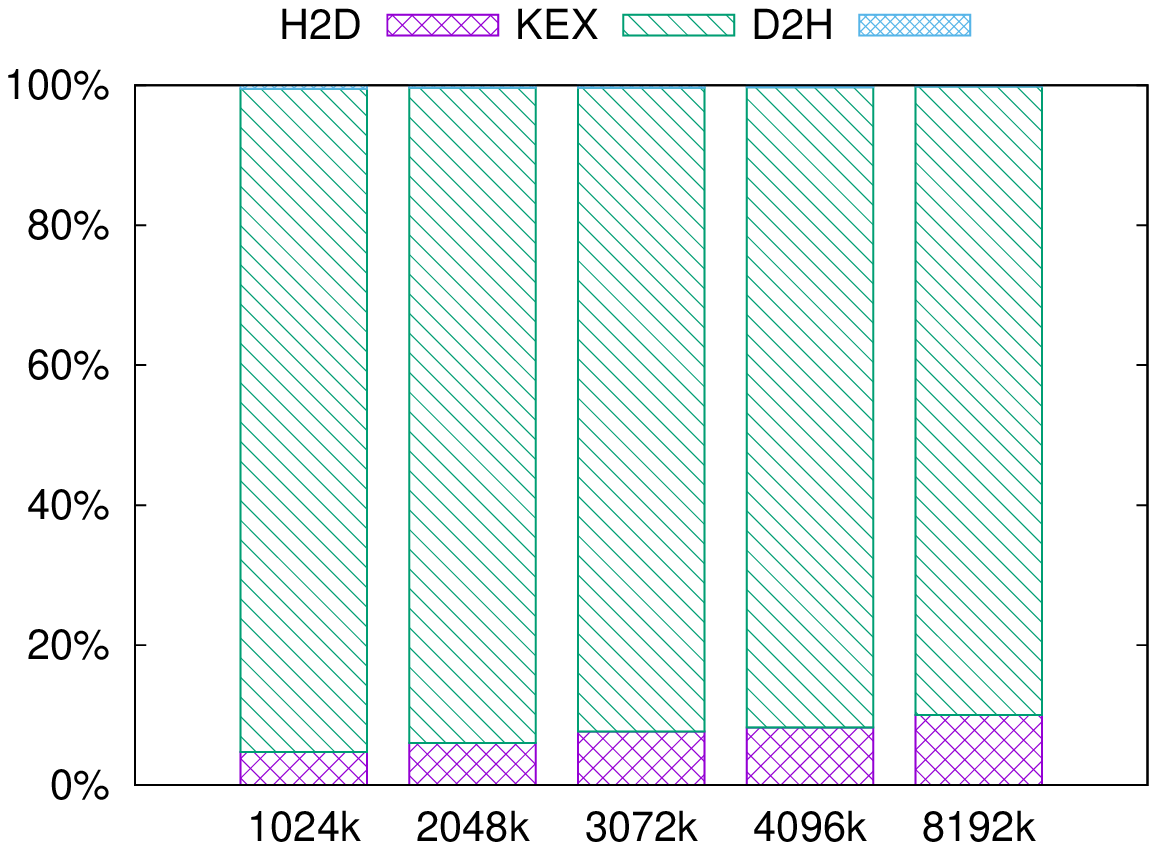}}
\subfigure[reduction v2.]{\label{fig:stat:reductionv2}\includegraphics[width=0.35\textwidth]{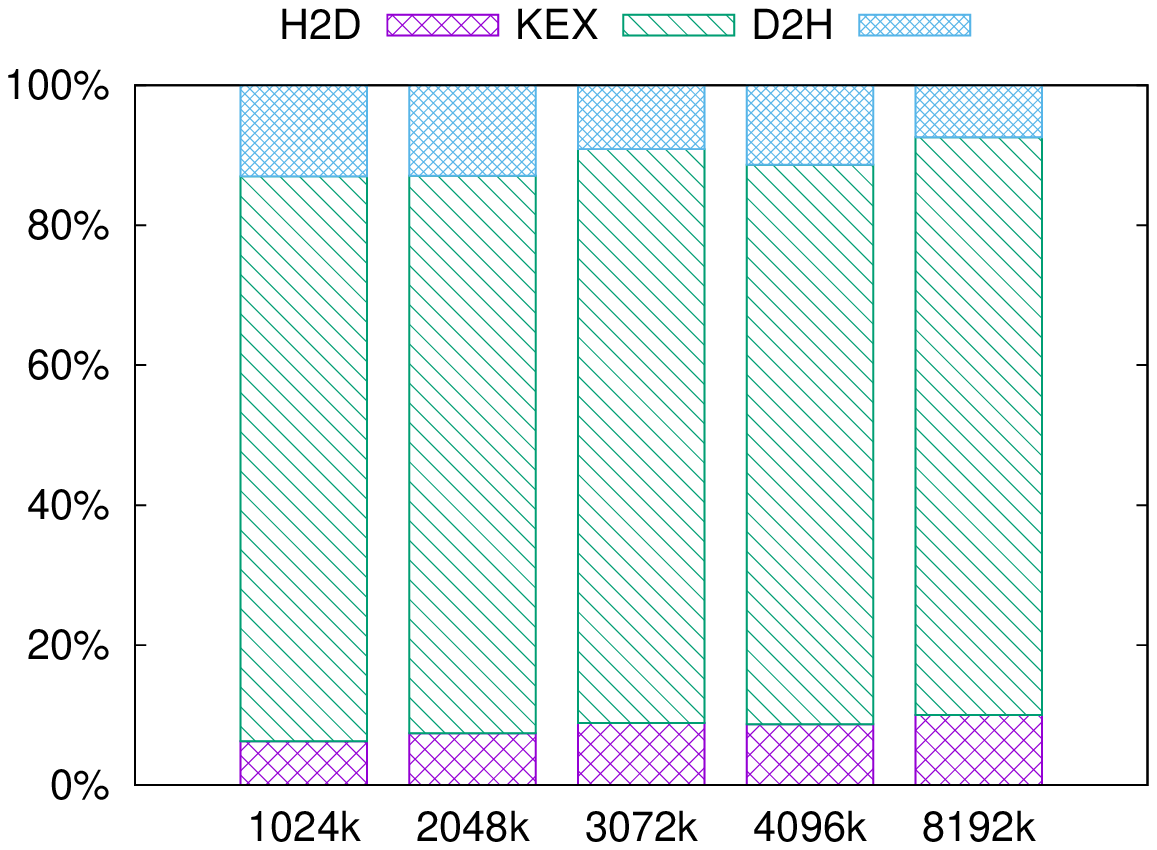}}
\caption{\texttt{R} changes over code variants of \texttt{NVIDIA Reduction}. }
\label{fig:stat_code_variant}
\end{figure}

Figure~\ref{fig:stat_code_variant} shows how data transfers change with the two code variants for \texttt{Reduction}. \texttt{Reduction v1}  performs the whole reduction work on the accelerator, thus significantly reducing the data-moving overheads. Meanwhile, \texttt{Reduction v2}  performs the final reduction on the host side, and thus needs to transfer the intermediate results back to host. Therefore, different code variants will generate different data transferring requirements, which is to be taken into account when streaming such code variants. 

\subsubsection{The Impact of Platform Divergence} 
\begin{figure}[!h]
\centering
\subfigure[ on MIC.]{\label{fig:stat:platform:a}\includegraphics[width=0.35\textwidth]{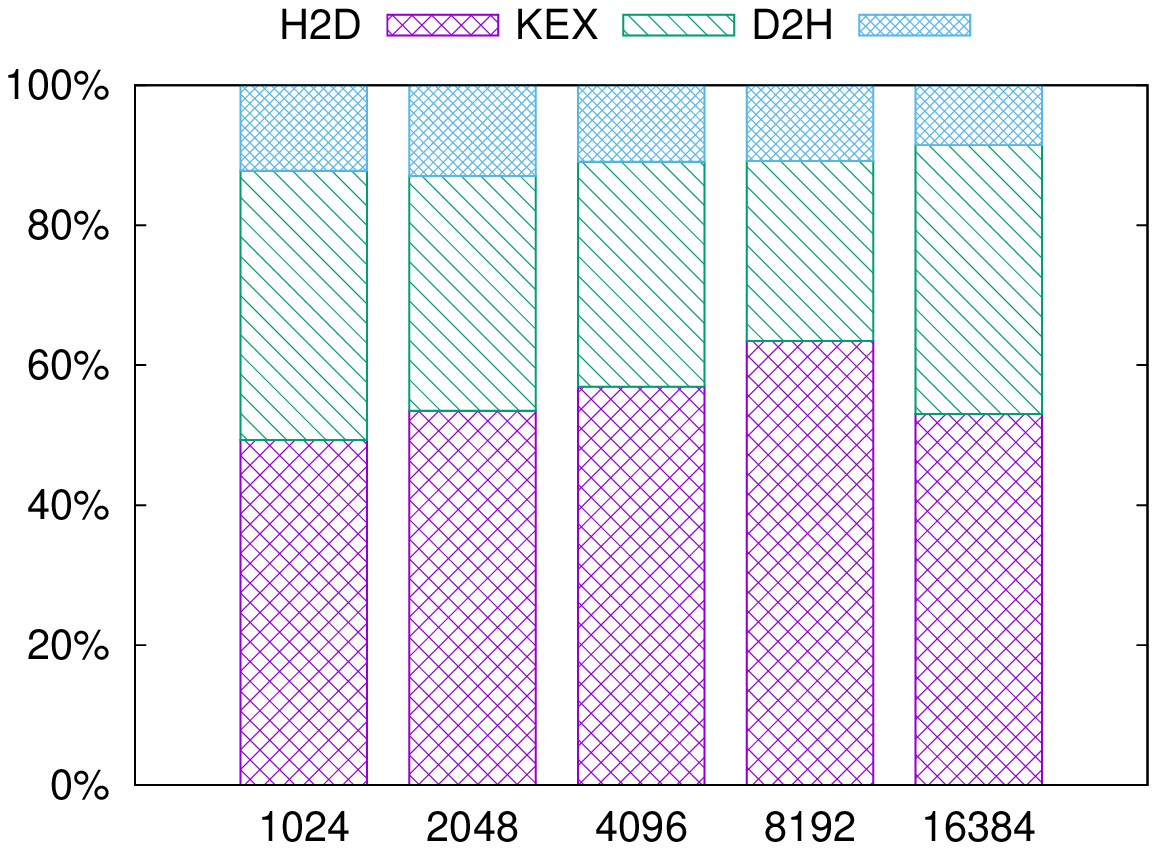}}
\subfigure[on GPU.]{\label{fig:stat:platform:b}\includegraphics[width=0.35\textwidth]{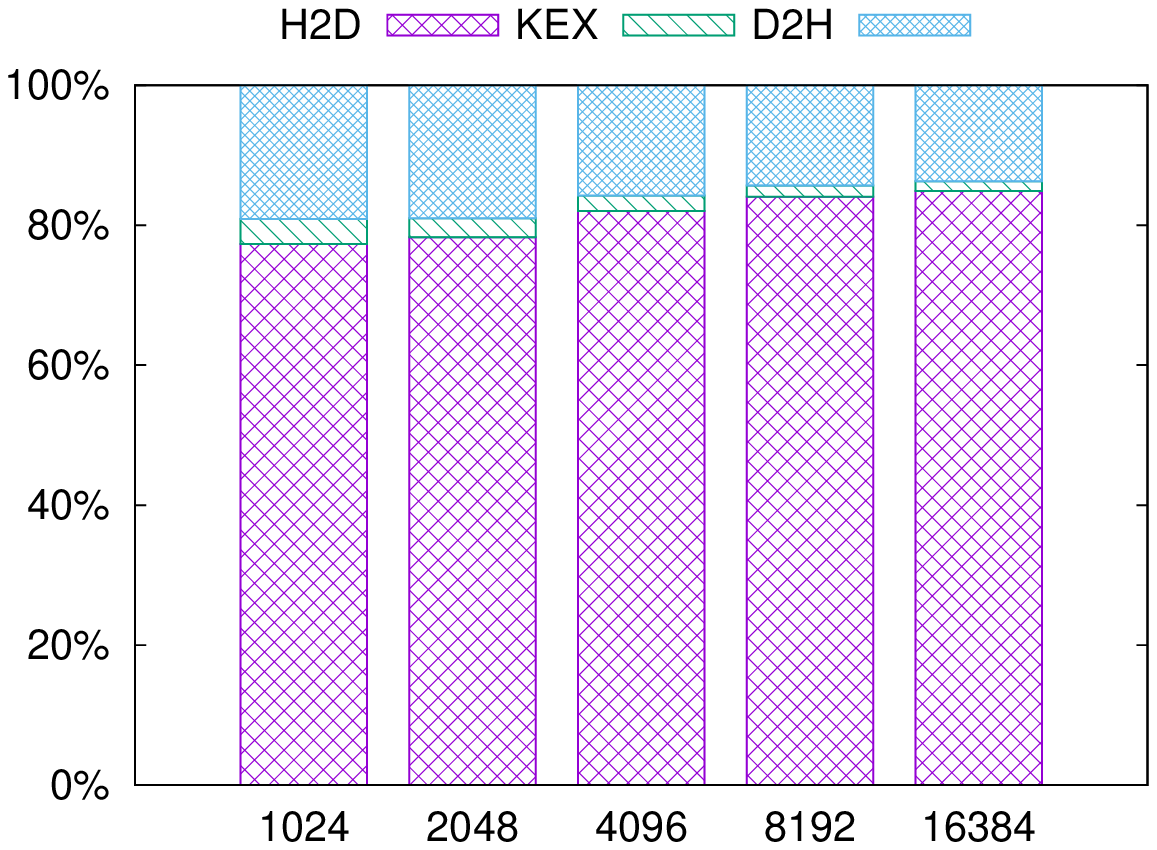}}
\caption{\texttt{R} changes over platforms of \texttt{Rodinia nn}. }
\label{fig:stat:platform}
\end{figure}

Figure~\ref{fig:stat:platform} shows how $R$ changes on MIC and a K80 GPU~\footnote{Note that the only difference lies in devices (Intel Xeon 31SP Phi versus NVIDIA K80 GPU) and all the other configurations are the same.}. We see that the kernel execution time (of \texttt{nn}) on the MIC occupies 33\% on average  while the number is only around 2\% on the GPU. This is due to the huge processing power from NVIDIA K80, which reduces the \texttt{KEX} fraction significantly. Ideally, using the streaming mechanism can improve the overall performance by 2\% on the GPU. In this case, we argue that it is unnecessary to use multiple streams on GPU. 

%\subsubsection{Discussion}
To summarize, we observe that $R$ varies over platforms, benchmarks, code variants, and input configurations. Each benchmark has a unique balance between computation and memory accesses. Different code variants lead to the differences in transferred data amounts and kernel execution time. Also, input configurations can incur changes in transferred data amounts and/or kernel execution time. Furthermore, $R$ depends on hardware capabilities (e.g., the PCIe interconnect and the accelerator).  

We use $R$ as an indicator of deciding whether the target application is worthwhile to be streamed.  Figure~\ref{fig:cdf} shows that \texttt{H2D} takes only 10\% of the total execution time for over 50\% test cases. We argue that, on the one hand, the applications are not worthwhile to be streamed when $R$ is small. This is due to two factors: (1) code streaming introduces overheads for filling and emptying the pipeline, and (2) streaming an application requires extra programming efforts from reconstructing data structures and managing streams.  Thus, streaming such applications might lead to a performance degradation compared with the non-streamed code. On the other hand, when $R$ is too large (e.g., 90\%), it is equally not worthwhile to apply streams. When the fraction of \texttt{H2D} is too large, using accelerators may lead to a performance drop (when comparing to the case of only using CPUs), not to mention using streams. In real-world cases, users need make the streaming decision based on the value of $R$ and the coding effort. 

\section{Our Streaming Approach} \label{sec:approach}
%\subsection{Analyzing Codes}
%For the applications, we need to state whether they can be pipelined. If not, why? If yes, tell us how to in the next subsection. 
%Given an heterogeneous code, we use the following steps to  determine whether it can be streamed. 

\subsection{Categorization}
%For each application, we need to determine whether it can be pipelined. We sum up to three steps: (1) get the data transfer ratio R (the fraction of the data transfer time to the total execution time): run the application codes in a strictly stage-by-stage manner, and get the time of data transmission and kernel execution, respectively, and then get the ratio R; (2) if R is less than 10\% or greater than 90\%, we conclude that the application is not suitable for multiple streams; otherwise we still need the third step; (3) judge whether the application is overlappable or not. Next we will detailedly introduce the third step.

\begin{figure}[!h]
\centering
\subfigure[]{\label{fig:code:flow0}\includegraphics[width=0.24\textwidth]{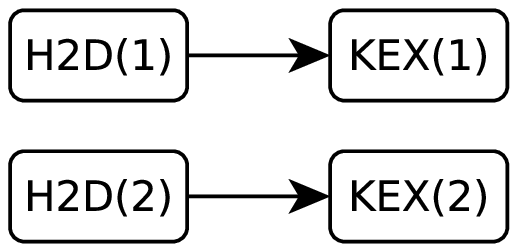}}
\subfigure[]{\label{fig:code:flow1}\includegraphics[width=0.35\textwidth]{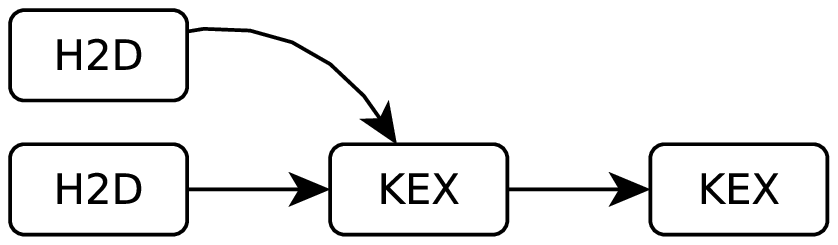}}
\subfigure[]{\label{fig:code:flow2}\includegraphics[width=0.35\textwidth]{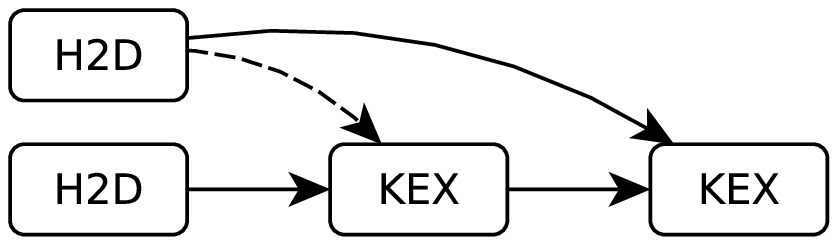}}
\caption{The dependent relationship between \texttt{H2D} and \texttt{KEX} (The number in the parenthesis represents stages from different tasks). } 
\label{fig:code:flow}
\end{figure}

After determining the necessity to apply the pipelining/streaming mechanism, we further investigate \textit{how-to}. Generally, we divide applications into tasks which are mapped onto different processing cores. As we have mentioned above, each task includes the subtasks of data transfers and kernel execution. To pipeline codes, we should guarantee that \textit{there exist independent tasks} running concurrently. 
%Multiple streams improve performance by hiding data transfer time. Say concretely, we must divide the application into small subtasks, and then by pipelining mechanism could overlap data transfer and kernel execution. So for the application, we must judge whether it could be divided into small subtasks. In our experiments, we find that the benchmark, which have iterative operation in the kernel function, frequently belongs to the one non-overlappable. In this case, we just need one data transmission operation, but in the device sink, we calculate the same input data iteratively with no interaction between host and device. If we compulsively overlap the two operation, we will get extra transmission overhead.
% How-to
%Independent tasks are referred to be as the tasks that can run simultaneously. 
%There being independent tasks (). 
Once discovering independent tasks, we are able to overlap the execution of \texttt{H2D} from one task and \texttt{KEX} from another (Figure~\ref{fig:code:flow0}). For a single task, \texttt{H2D} is dependent on \texttt{KEX}. 
%In the following, we will discover task concurrency by code analysis. 
% shows the \texttt{H2D} and the \texttt{KEX} step for a single task, and \texttt{H2D} is dependent on \texttt{KEX}.  

In practice, more than one \texttt{H2D} may depend on a single kernel execution (Figure~\ref{fig:code:flow1}). Thus, we need to analyze each \texttt{H2D}--\texttt{KEX} dependency to determine whether each pair can be overlapped. Moreover, an application often has more than one kernel. Implicitly, each kernel is synchronized at the end of its execution. Therefore, the kernel execution order is strictly respected within a single task. Figure~\ref{fig:code:flow2} shows that \texttt{H2D(1)} is depended by \texttt{KEX(1)}, but the data is not used til the execution of \texttt{KEX(2)}. Thus, this data transfer can be delayed right before \texttt{KEX(2)} when analyzing dependency and/or streaming the code. 

Based on the dependency analysis, we categorize the codes listed in Table~\ref{tbl:benchmark_intro} as \textit{streamable codes} (Section~\ref{sec:streaming}) and \textit{non-streamable codes}. 
%When analyzing the codes , we summarize that several patterns of the heterogeneous codes cannot be streamed. 
The first pattern (\texttt{SYNC}) of \textit{non-streamed codes} is when the \texttt{H2D} data is shared by all the tasks of an application. 
%In this case, the whole data transfer has to be finished before kernel execution (Figure~\ref{fig:code:nonstreamable0}).  Figure~\ref{fig:code:nonstreamable1} shows the second non-streamable pattern (\texttt{Iterative}). We see that the kernel execution will be invoked in an iterative manner once the data is located on device. Although such cases can be streamed by overlapping the data transfer and the first iteration of kernel execution, we argue that the overlapping will not be beneficial for a large number of iterations. 
In this case, the whole data transfer has to be finished before kernel execution.  The second non-streamable pattern is characterized as \texttt{Iterative}, for which \texttt{KEX} will be invoked in an iterative manner once the data is located on device. Although such cases can be streamed by overlapping the data transfer and the first iteration of kernel execution, we argue that the overlapping brings no performance benefit for a large number of iterations.

%\begin{figure}[!h]
%\centering
%\subfigure[SYNC]{\label{fig:code:nonstreamable0}\includegraphics[width=0.24\textwidth]{dia/code_nonstream_0.eps}}
%\subfigure[Iterative]{\label{fig:code:nonstreamable1}\includegraphics[width=0.30\textwidth]{dia/code_nonstream_1.eps}}
%\caption{Patterns of nonstreamable codes  (The number in the parenthesis represents stages from different streams). } 
%\label{fig:code:nostreamable}
%\end{figure}

% results
We analyze the heterogeneous codes listed in Table~\ref{tbl:benchmark_intro} and categorize them in Table~\ref{tbl:benchmark_category}. Each nonstreamable code is labeled with \texttt{SYNC} or \texttt{Iterative}. We note that the kernel of \texttt{hearwall}  has such a large number of lines of codes and complex structures that its execution takes a major proportion of the end-to-end execution time. It is unnecessary to stream such code on any platform. Due to the multiple \texttt{H2D}--\texttt{KEX} dependency pairs, an application might fall into more than one category (e.g., \texttt{streamcluster}). Also, the kernel of \texttt{myocyte} runs sequentially and thus there are no concurrent tasks for the purpose of pipelining. For the streamable codes, we group them into three categories, which are detailed in Section~\ref{sec:streaming}. 

%Now we could get judgement about whether one application is overlappable. If the application is overlappable, we could streamlize it according to the methods which are explained in the next section 4.2. One caveat, however, is that even for one overlappable application we could probably not get performance improvement by multiple streams mechanism due to extra overhead, we will provide examples int section 4.2.

\begin{table*}[!t]
\caption{Application Categorization.}
\begin{center}
\scalebox{0.70}{
\begin{tabular}{|c|c|c|c|c|c|}
\hline
&  \multicolumn{2}{|c|}{\textbf{Nonstreamable}}   & \multicolumn{3}{|c|}{\textbf{Streamable}}  \\ \hline 
& \textbf{SYNC} & \textbf{Iterative}  & \textbf{Independent} & \textbf{False-dependent} & \textbf{True-dependent} \\ \hline
%%%%%%%%%%%%%%%%%%%%%%%%%%%%%%%%%%%% Rodinia
\textbf{Rodinia} &
% 1
\begin{tabular}[c]{@{}c@{}}
	backprop, bfs,\\ 
	b+tree,  Kmeans, \\ 
	streamcluster \\
\end{tabular}  & 
% 2
\begin{tabular}[c]{@{}c@{}}
	hotspot, \\
	pathfinder \\
\end{tabular}  & 

%4 
\begin{tabular}[c]{@{}c@{}}
	backprop, dwt2d,\\ 
	nn, 	srad, \\
	streamcluster\\
\end{tabular}  & 
% 5
\begin{tabular}[c]{@{}c@{}}
	lavaMD, \\ 
	leukocyte \\
\end{tabular}  &  
%6
\begin{tabular}[c]{@{}c@{}}
	gaussian, \\ 
	lud \\
	nw \\
\end{tabular}   \\ \hline

%%%%%%%%%%%%%%%%%%%%%%%%%%%%%%%%%%%% Parboil
\textbf{Parboil} &
% 1
\begin{tabular}[c]{@{}c@{}}
	spmv, \\ 
	tpacf, \\
	bfs, \\
	mri-q \\
\end{tabular}  & 
% 2
\begin{tabular}[c]{@{}c@{}}
	mri-gridding, \\ 
	cutcp \\	
\end{tabular}  & 

%4 
\begin{tabular}[c]{@{}c@{}}
	sgemm\\ 
\end{tabular}  & 
% 5
\begin{tabular}[c]{@{}c@{}}
	stencil, \\ 
	lbm \\
\end{tabular}  &  
%6
\begin{tabular}[c]{@{}c@{}}

\end{tabular}   \\ \hline

%%%%%%%%%%%%%%%%%%%%%%%%%%%%%%%%%%%% NVIDIA SDK
\textbf{NVIDIA SDK} &
% 1
\begin{tabular}[c]{@{}c@{}}
	Reduction-2, \\

\end{tabular}  & 
% 2
\begin{tabular}[c]{@{}c@{}}
	FDTD3d, \\ 
	DXTCompression \\
\end{tabular}  & 

%4 
\begin{tabular}[c]{@{}c@{}}
	BlackScholes, 	DCT8x8, \\ 
	DotProduct, 	Histogram, \\ 
	MatrixMul, 	MatVecMul, \\ 
	QuansirandomGenerator, \\ 
	Reduction, 	Transpose, \\ 
	Tridiagonal, VectorAdd \\ 
\end{tabular}  & 
% 5
\begin{tabular}[c]{@{}c@{}}
	ConvolutionSeparable \\ 
	FastWalshTransform \\
	ConvolutionFFT2D\\
\end{tabular}  &  
%6
\begin{tabular}[c]{@{}c@{}}

\end{tabular}   \\ \hline

%%%%%%%%%%%%%%%%%%%%%%%%%%%%%%%%%%%% AMD SDK
\textbf{AMD SDK} &
% 1
\begin{tabular}[c]{@{}c@{}}
	\\ 

\end{tabular}  & 
% 2
\begin{tabular}[c]{@{}c@{}}
	BitonicSort, \\ 
	FloydWarshall, \\ 
	RadixSort \\ 

\end{tabular}  & 

%4 
\begin{tabular}[c]{@{}c@{}}
	BinomialOption, \\ 
	MonteCarloAsian, \\ 
	RecursiveGaussian, \\ 
	ScanLargeArrays, 	\\
	URNG, PrefixSum \\

\end{tabular}  & 
% 5
\begin{tabular}[c]{@{}c@{}}
	BoxFilter, \\ 
	DwtHaar1D, \\ 
	StringSearch \\ 
\end{tabular}  &  
%6
\begin{tabular}[c]{@{}c@{}}

\end{tabular}   \\ \hline

\end{tabular}}
\end{center}
\label{tbl:benchmark_category}
\end{table*}

\subsection{Code Streaming} \label{sec:streaming} 
%For overlappable applications, on the whole, we divide the streamlization operation into three steps: (1) analyze the input and output data, judge whether there is dependence or not, (2) divide task into some subtasks, and correspondingly partition the input data, and (3) launch multiple streams, and map subtasks to different streams. When mapping tasks to streams, we can employ two strategies: circular mapping, and uniform mapping, as shown in figure~\ref{fig:CircularMapping} and figure~\ref{fig:UniformMapping}.
%
%\begin{figure}[!h]
%\centering
%\includegraphics[width=0.9\textwidth]{circular.eps}
%\caption{Circular Mapping}
%\label{fig:CircularMapping}
%\end{figure}
%\begin{figure}[!h]
%\centering
%\includegraphics[width=0.9\textwidth]{uniform.eps}
%\caption{Uniform Mapping}
%\label{fig:UniformMapping}
%\end{figure}

We divide the streamable/overlappable applications into three categories based on task dependency: (1) embarrassingly independent, (2) false dependent, and (3) true dependent. Tasks are generated based on input or output data partitioning, and thus task dependency shows as a form of data dependency. We will explain them one by one.  

%Different applications own diverse features. When programming, we can classify into three categories on the basis of the dependencies between tasks: data independent, data dependent \& data read-only, and data dependent \& data read-write. Next section explains the three patterns and how to streamlize.

%\subsubsection{Data Independent}
\subsubsection{Embarrassingly Independent} 
\begin{figure}[!h]
\centering
\includegraphics[width=0.70\textwidth]{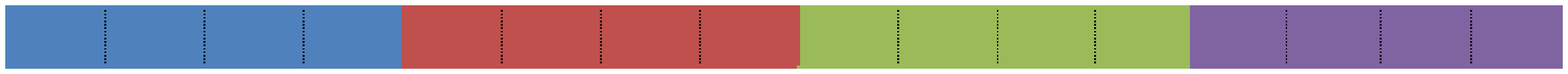}
\caption{Nearest Neighbor Data Partition (16 elements and 4 groups).}
\label{fig:nn}
\end{figure}

Tasks from such overlappable applications are completely independent. Thus, there is no data dependency between tasks. Taking \texttt{nn} (nearest neighbor) for example, it finds the k-nearest neighbors from an unstructured data set. The sequential \texttt{nn} algorithm reads in one record at a time, calculates the Euclidean distance from the target latitude and longitude, and evaluates the k nearest neighbors. By analyzing its code, we notice no dependency between the input data. Figure~\ref{fig:nn} shows how to partition the input data. Assuming 16 elements in the set, we divide them into 4 groups, which represent 4 tasks. Then we spawn streams to run the tasks. Due to no dependency, data transferring from one task can overlap kernel execution from another. More \texttt{Embarrassingly Independent} applications are shown in Table~\ref{tbl:benchmark_category}. 

%The simplest case for streamlizing overlappable applications is data independence which we can block the data directly. NN, nearest neighbor finds the k-nearest neighbors from an unstructured data set. The sequential NN algorithm reads in one record at a time, calculates the Euclidean distance from the target latitude and longitude, and evaluates the k nearest neighbors. Analyzing the program logic relationship, there is no dependencies between the input data. Figure~\ref{fig:nn} shows how to block the input dataset. Assuming 16 elements in the set, we divides them into 4 blocks, corresponds to 4 subtasks. Now we launch multiple streams and adopt above mapping strategies to carry out tasks.

%\subsubsection{Data Dependent \& Data Read-only}
\subsubsection{False Dependent} 
\begin{figure}[!h]
\centering
\subfigure[Input elements dependencies.]{\label{fig:fwt1}\includegraphics[width=0.70\textwidth]{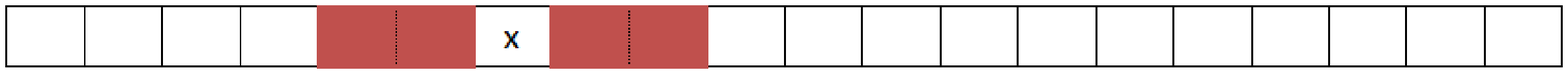}} \\
\subfigure[Data Partition]{\label{fig:fwt2}\includegraphics[width=0.70\textwidth]{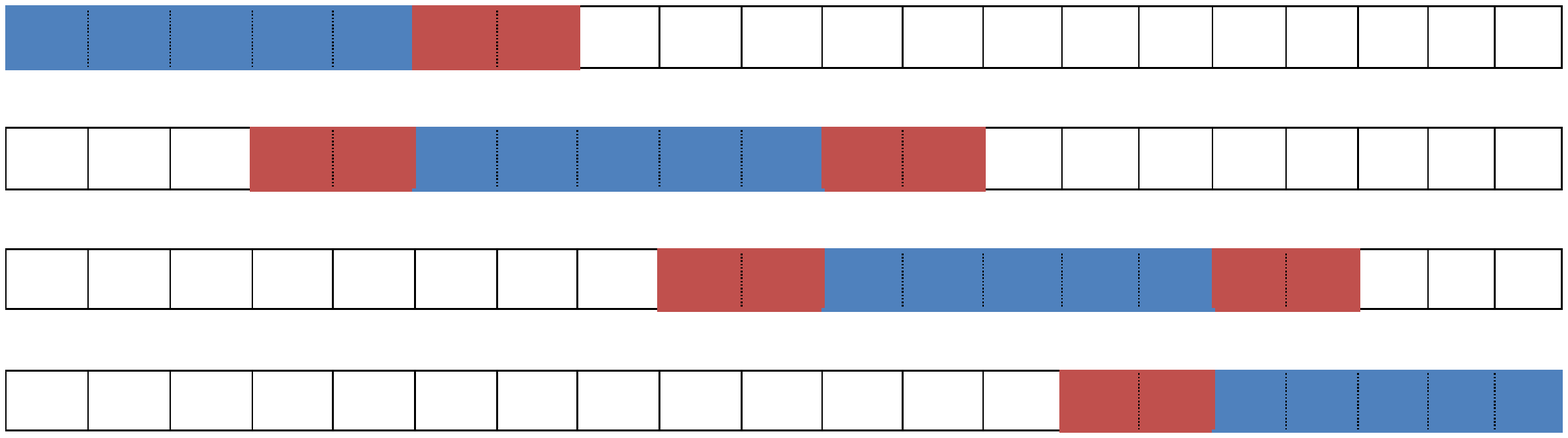}}
\caption{Task dependency and data partition for \texttt{FWT}}
\label{fig:fwt}
\end{figure}

There exist data dependencies in such overlappable applications, but the dependencies come from read-only data (i.e., \textit{RAR} dependency).  In this case, two tasks will share common data elements. A straightforward solution to this issue is that each task moves the shared data elements separately. For example,  \texttt{FWT} (fast walsh transform) is a class of generalized Fourier transformations. By analyzing the code, we find that there are dependencies between the input data elements: As shown in Figure~\ref{fig:fwt1}, calculating element $x$ is related to the 4 neighbors which are marked in red. The input elements are read-only, so we can eliminate the relationship by redundantly transferring boundary elements (Figure~\ref{fig:fwt2}). We first divide the total elements into four blocks, corresponding to four tasks (in blue). Then, we additionally transfer the related boundary elements (in red) when dealing with each data block. More \texttt{False Dependent} applications are shown in Table~\ref{tbl:benchmark_category}. 
%The second pattern for streamlizing overlappable applications is data dependent \& data read-only. In this case, as long as the dependency has rules to follow, the dependency can be easily processed. FWT, fast walsh transform, belong to a class of generalized Fourier transformations. They have applications in various fields of electrical engineering and numeric theory. First analyzing the code, there is dependency between input data elements: As shown in Figure~\ref{fig:fwt1}, element x is related to 4 neighbors which are marked in red, that is when calculating element x we must first get the 4 related neighbors. The input elements is read-only, so we can easily solute the relationship by multiple transferring boundary relevant values. Figure~\ref{fig:fwt2} shows how we get rid of data dependency. First of all, we divide the total elements into four blocks, corresponding to four subtasks. Second, when transferring every data block, we must additionally transfer the related boundary value as shown in red, in this way, solving cleverly the dependencies among boundary elements. So we can easily launch multiple streams and use the above mapping strategies to handle subtasks concurrently. 

\subsubsection{True Dependent}

The third category of overlappable applications is similar to the second one in that there exist data dependencies between tasks. The difference is that the dependency is true (i.e., \textit{RAW}). This is complicated for programmers  not only because there is a dependence between the input data elements, but because they need to update input data in the process of calculation. Thus, the output elements depend on the updated input data, and we must control the order of calculation. The key for this pattern is to discover concurrency while respecting the dependency.

%The third pattern for streamlizing overlappable applications is data dependent \& data read-write, which is the most complicated situation for programmers. The situation is more complex, there is dependence between the input data elements, and need to update input data in the process of calculation, the output elements depend on the updated input data, so we must control the order of calculation. The key for this pattern is to clarify dependencies, and then finds the codes which can be executed concurrently.

\begin{figure*}[!h]
\centering
\subfigure[Dependencies.]{\label{fig:nw1:a}\includegraphics[width=0.24\textwidth]{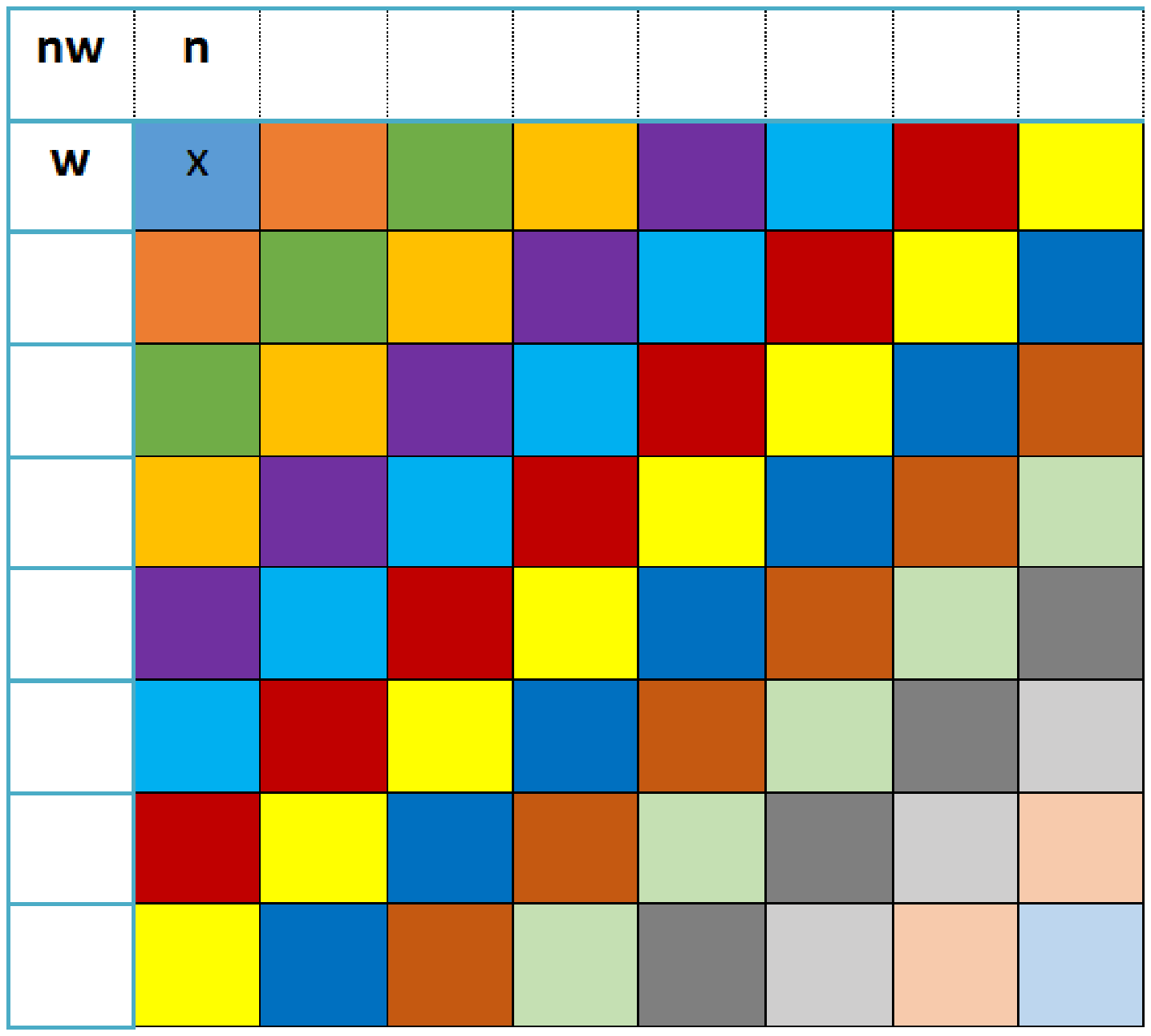}}
\subfigure[Block Number.]{\label{fig:nw1:b}\includegraphics[width=0.36\textwidth]{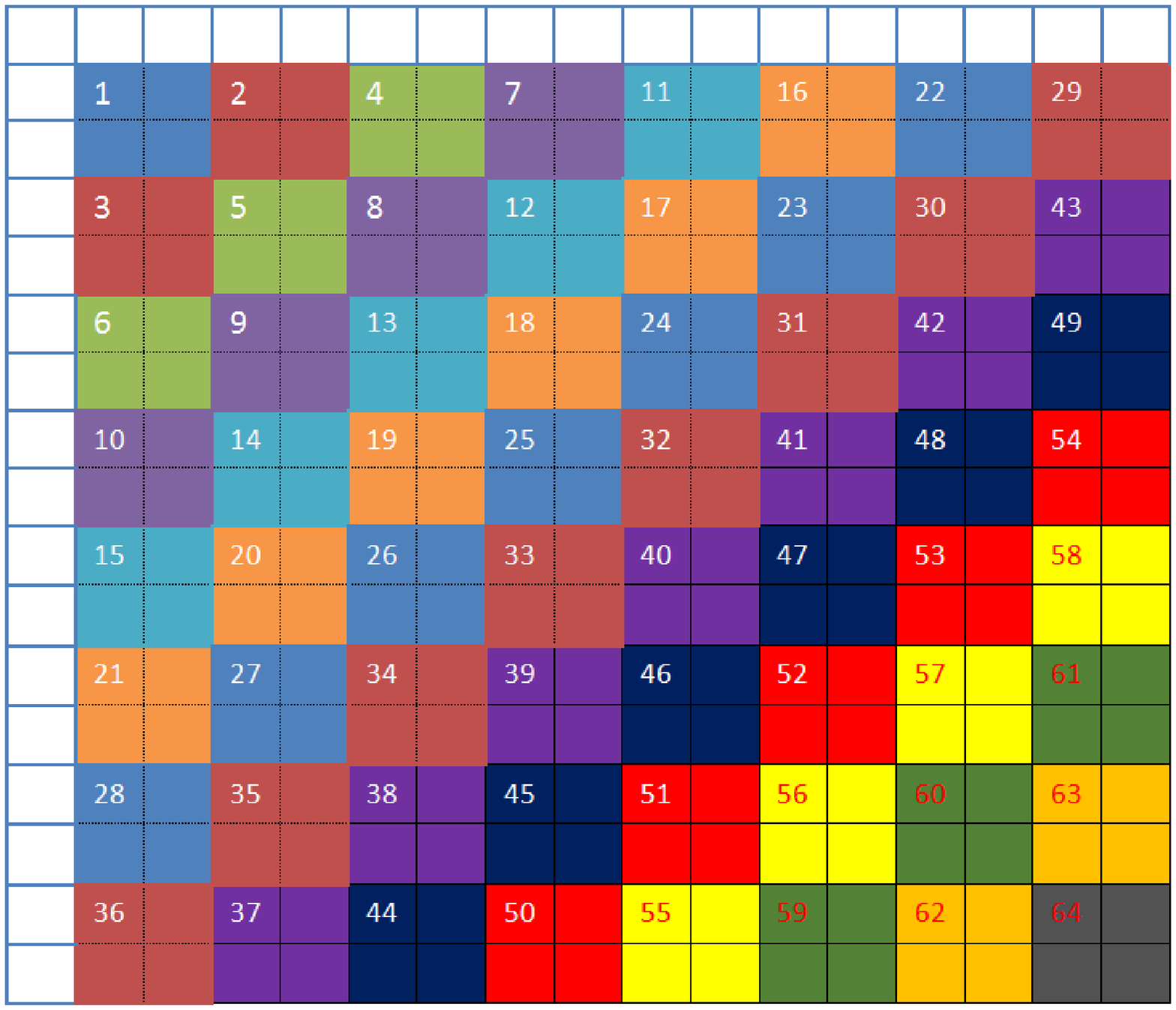}}
\subfigure[Storage Pattern.]{\label{fig:nw1:c}\includegraphics[width=0.36\textwidth]{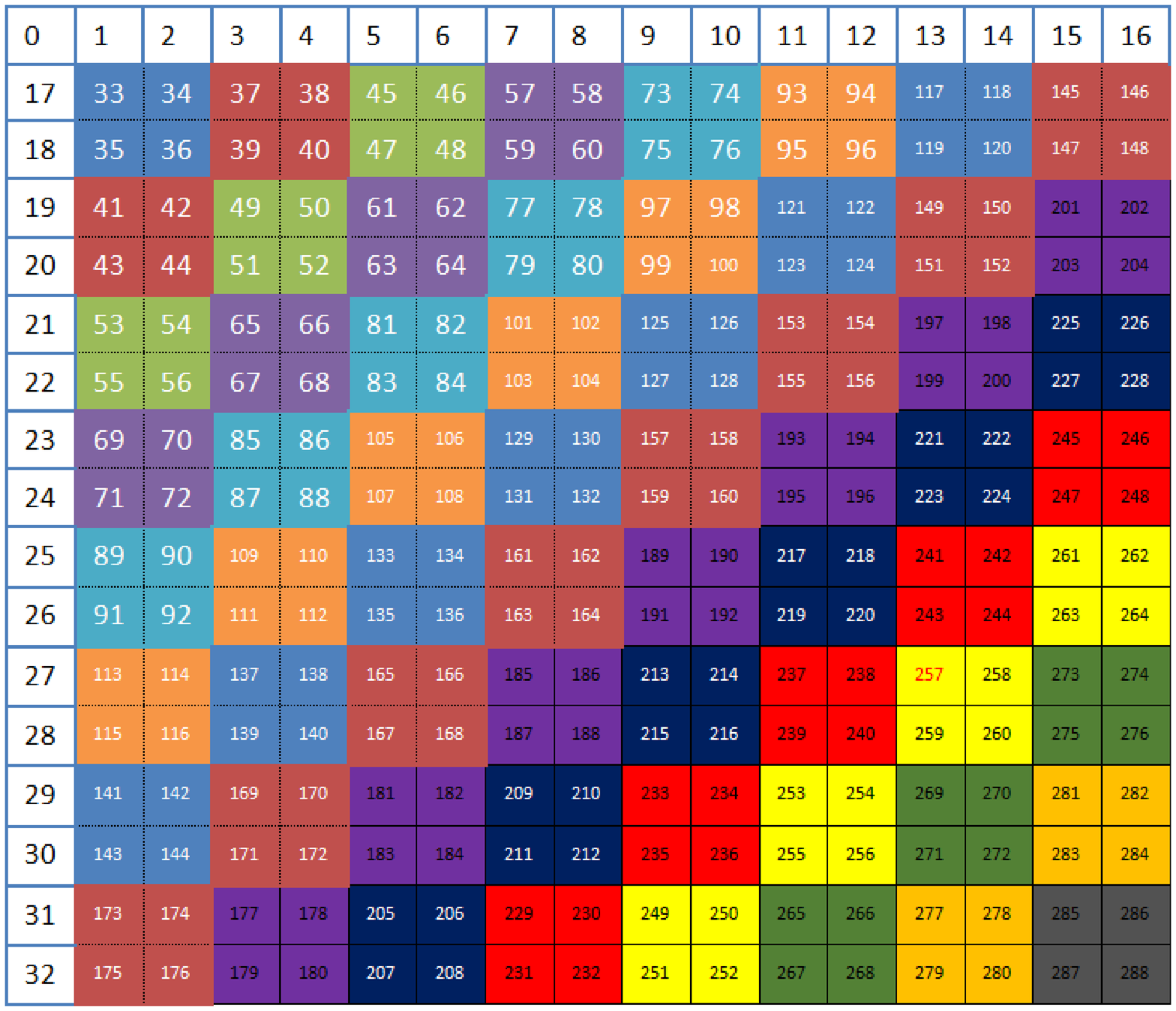}}
\caption{\texttt{NW} input elements dependencies and how to partition.}
\label{fig:nw1}
\end{figure*}

\texttt{NW}, Needleman-Wunsch is a nonlinear global optimization method for DNA sequence alignments. The potential pairs of sequences are organized in a 2D matrix. In the first step, the algorithm fills the matrix from top left to bottom right, step-by-step. The optimum alignment is the pathway through the array with maximum score, where the score is the value of the maximum weighted path ending at that cell. Thus, the value of each data element depends on the values of its northwest-, north- and west-adjacent elements. In the second step, the maximum path is traced backward to deduce the optimal alignment. As shown in Figure~\ref{fig:nw1:a}, calculating element $x$ is related with three elements: `n' (north-element), `w' (west-element), and `nw' (northwest-element). We must calculate output elements diagonal by diagonal (in the same color), 
%according to the order of oblique diagonal marked by the same color, 
and the elements on the same diagonal can be executed concurrently. Figure~\ref{fig:nw1:b} shows how we divide the data: we number all blocks from the top-left diagonal to the bottom-right one (the first row and first column are the two DNA sequences, marked in number 0), and then change the storage location to let elements from the same block stored contiguously.
% and to make the blocks stored according to marked numbers. 
Figure~\ref{fig:nw1:c} shows the storage pattern, and the numbers represent the relative location. By controlling the execution in the order of diagonal from top-left to bottom-right, we can respect the dependencies between tasks. Further, the tasks on the same diagonal can run concurrently with multiple streams. Note that the number of streams changes on different diagonals. More \texttt{True Dependent} applications can be found in Table~\ref{tbl:benchmark_category}. 

%\begin{figure}[!h]
%\centering
%\includegraphics[width=1.0\textwidth]{nw3.eps}
%\caption{NW input elements relative storage locations}
%\label{fig:nw2}
%\end{figure}
%\begin{figure*}[!h]
%\centering
%\subfigure[LUD input elements dependencies.]{\includegraphics[width=0.4\textwidth]{lud1.eps}}
%\subfigure[Potential Parallel Region.]{\includegraphics[width=0.5\textwidth]{lud2.eps}}
%\caption{LUD input elements dependencies and how to partition.}
%\label{fig:lud}
%\end{figure*}

%LUD, LU Decomposition, is an algorithm to calculate the solutions of a set of linear equations. It is similar to NW, belong to the third patterns. Figure~\ref{fig:lud} shows the data dependencies and the parallel region. For (a), elements in the two color zones have different dependencies: for example, element y depends on y0 $\sim$ y6, element n on n0 $\sim$ n7. Based on this, we get the (b) potential parallel region: there is 15 region colored by fifteen different colors and marked by number, in addition, the marked number also represents the execution order, we will perform serially, carry out synchronous operation between two marked number, however, in the same number, we may launch multiple streams.

\section{Experimental Results} \label{sec:exp}
In this section, we discuss the performance impact of using multiple streams. We use the CPU-MIC heterogeneous platform detailed in Section~\ref{subsec:exp:platform}. Due to the limitation in time and space, we port 13 applications from Table~\ref{tbl:benchmark_intro} with \texttt{hStreams}. 
%As shown in Table~\ref{tbl:ApplicationsPatterns}, we use 13 benchmarks, which belong to different patterns, then map to the three patterns(\textrm{i}, \textrm{ii}, \textrm{iii} represents `Embarrassingly Independent', `False Independent' and `True Independent', respectively) to streamlize the application code.
As shown in Table~\ref{tbl:benchmark_category}, these 13 benchmarks are characterized as different categories and thus we use the corresponding approach to stream them.

\begin{figure}[!t]
\centering
\includegraphics[width=0.9\textwidth]{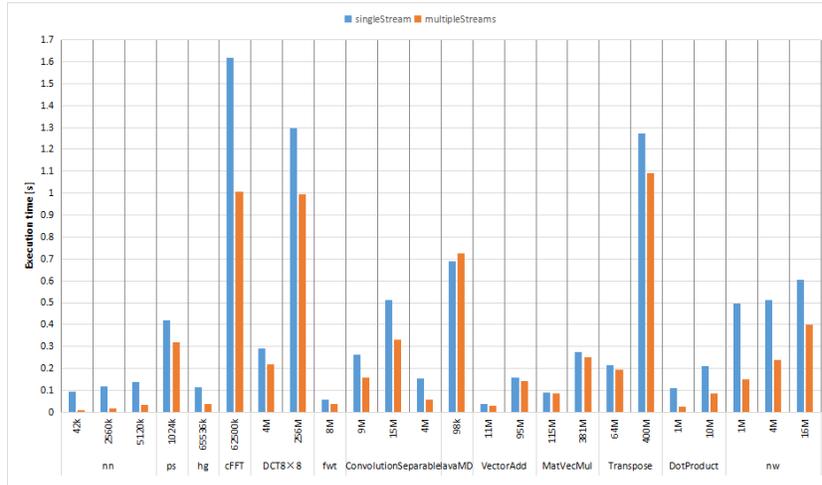}
\caption{A performance comparison between single stream and multiple streams. For each application, we employ different configuration, corresponding to different data size. Note that \texttt{ps}, \texttt{hg}, \texttt{cFFT} and \texttt{fwt} represents \texttt{PrefixSum}, \texttt{Histogram}, \texttt{ConvolutionFFT2D} and \texttt{FastWalshTransform}, respectively.} 
\label{fig:result}
\end{figure}

Figure~\ref{fig:result} shows the overall performance comparison. We see that using multiple streams outperforms using a single stream, with a  performance improvement of 8\%--90\%. In particular, for \texttt{nn}, \texttt{FastWalshTransform}, \texttt{ConvolutionFFT2D}, and \texttt{nw}, the improvement is around 85\%, 39\%, 38\%, and 52\%, respectively. However, for applications such as \texttt{lavaMD}, we cannot obtain the expected performance improvement with multiple streams, which will be discussed in the following. 
%, we will give a discussion in third later.

Also, we notice that the performance increase of using multiple streams varies over benchmarks and datasets. This is due to the differences in data transfer ratio ($R$): a larger $R$ leads to a greater performance improvement. For example, for \texttt{ConvolutionSeparable} and \texttt{Transpose}, the average performance improvement is 45\% and 11\%, with $R$ being 19\% and 14\%, respectively. Further, when selecting two datasets (400M and 64M) for \texttt{Transpose}, we can achieve a performance increase of 14\% and 8\%, with $R$ being 20\% and 10\%, respectively. 

%for different benchmarks or for the same benchmark but different data size, the scope of performance improvement is discriminating, that is because of diverse data transfer ratio(R), when R is in the reasonable range(e.g.,10\%--90\%), the more the value of R, the greater the performance improvement. For example, for ConvolutionSeparable and Transpose, the average performance improvement is 45\% and 11\%, with average value of R 19\% and 14\%. Otherwise, for benchmark Transpose, choosing two problem size: 400M and 64M, we also get diverse performance improvement: 14\% and 8\%, also different R: 20\% and 10\%.

%\begin{table*}[!h]
%\caption{Applications and Patterns}
%\begin{center}
%\scalebox{0.65}{
%\begin{tabular}{|c|c|c|c|c|c|}
%\hline
%Applications & Pattern & Applications & Pattern & Applications & Pattern \\
%\hline
%nn & \textrm{i} & DotProduct & \textrm{i} & VectorAdd & \textrm{i} \\
%\hline
%MatVecMul & \textrm{i} & Transpose & \textrm{i} & DCT8x8 & \textrm{i} \\
%\hline
%PrefixSum & \textrm{i} & Histogram & \textrm{i} & ConvolutionFFT2D & \textrm{ii} \\
%\hline
%FastWalshTransform & \textrm{ii} & ConvolutionSeparable, lavaMD & \textrm{ii} & nw & \textrm{iii} \\
%\hline
%\end{tabular}}
%\end{center}
%\label{tbl:ApplicationsPatterns}
%\end{table*}

For the \texttt{False Dependent} applications, if the extra overhead of transferring boundary elements is nonnegligible, code streaming is not beneficial. For \texttt{FWT}, one element is related to 254 elements which is far less than the subtask data size of 1048576. Therefore, although having to transfer extra boundary values, the overall streaming performance impact is positive. However, when the boundary elements are almost equal to the subtask size, the overhead introduced by boundary transmission can not be ignored. \texttt{LavaMD} calculates particle potential and relocation due to mutual forces between particles within a large 3D space. In the experiment, one element for lavaMD depends on 222 elements, in which 111 elements are lying before the target element and the other half behind. The task data size is 250, which is close to the boundary element number. Thus, we cannot get the expected performance improvement, and the experimental results confirm our conclusion. Specifically, when the task is of 250 and remains unchanged, for single stream, the \texttt{H2D} and \texttt{KEX} time is 0.3476s and 0.3380s, respectively. When using multiple streams, the overall execution time is 0.7242s. Therefore, it is not beneficial to stream the overlappable applications like \texttt{lavaMD}. 

%\section{Experimental Results} \label{sec:exp}
%\subsection{Using Scenario} \label{sec:flow}
\section{Conclusion}
In this paper, we summarize a systematic approach to facilitate programmers to determine whether the application is required and worthwhile to use streaming mechanism, and how to stream the code. (1) obtaining the ratio $R$: run the codes in stage-by-stage manner, record the \texttt{H2D} and \texttt{KEX} time, and  calculate $R$; (2) judging whether the application is overlappable;  (3) streaming the codes by either eliminating or respecting data dependency. Our experimental results on 13 streamed benchmarks show a performance improvement of upto 90\%. 
% if R is between 10\% and 90\%, goto next step, else get conclusion that application is not suitable for streaming; (2) judge whether the application is overlappable: if the kernel function possess iteration, then it is not streamlizable; (3) analyse codes, ravel out the dependency, according to the above three patterns: data independent, data dependent \& data read-only, data dependent \& data read-write, streamlize the application.

The process of analyzing whether a code is streamable and transforming the code is manually performed. Thus, we plan to develop a compiler analysis and tuning framework to automate this effort. Based on the streamed code, we will further investigate how to get optimal performance by setting a proper task and/or resource granularity. Ultimately, we plan to autotune these parameters leveraging machine learning techniques. Also, we want to investigate the streaming mechanism on more heterogeneous platforms, other than the CPU-MIC one. 

%Now we get streamlization methods, but it does not mean that we could get the optimal performance, because task and resource granularity affect the performance, we must determine the optimal parameters values. In the future, we would like to leverage machine learning techniques to obtain a proper task and resource granularity.

\section*{Acknowledgment}

\small
We would like to thank the reviewers for their constructive comments. This work was partially funded by the National Natural Science Foundation of China under Grant No.61402488 and No.61502514, the National High-tech R\&D Program of China (863 Program) under Grant No. 2015AA01A301, the National Research Fundation for the Doctoral Program of Higher Education of China (RFDP) under Grant No. 20134307120035 and No. 20134307120031. 

{\small
\bibliographystyle{plain}
\bibliography{mybib}  % sigproc.bib is the name of the Bibliography in this case
}

\end{document}